\documentclass[aps,pra,twocolumn,superscriptaddress, nofootinbib,longbibliography]{revtex4-1}
\pdfoutput=1
\usepackage[utf8]{inputenc}
\usepackage{bbm}
\usepackage[english]{babel}
\usepackage[T1]{fontenc}
\usepackage{amsmath}
\usepackage{placeins}
\usepackage{hyperref}
\usepackage{tikz}
\usepackage{lipsum}
\usepackage{float}
\usepackage{graphicx}
\DeclareUnicodeCharacter{2212}{\textminus}
\usepackage{latexsym}

\usepackage{amssymb}
\usepackage{csquotes}
\usepackage{nicefrac}
\usepackage{amsfonts}
\usepackage{enumerate}
\usepackage{upgreek}
\usepackage{subfigure}
\usepackage{bm}
\usepackage[version=3]{mhchem}
\usepackage{nicefrac}
\usepackage{bbold}
\usepackage{verbatim}
\usepackage{multirow}
\usepackage{color}
\usepackage{comment}
\usepackage{xcolor}
\usepackage{soul}

\usepackage{braket}

\newcommand{\C}{\hat{c}}
\newcommand{\Cd}{\hat{c}^\dagger}

\newcommand{\Bd}{\hat{b}^\dagger}
\newcommand{\B}{\hat{b}}

\newcommand{\n}{\hat{n}}
\newcommand{\Ham}{\hat{\mathcal{H}}}
\newcommand{\hc}{\mathrm{h.c.}}

\renewcommand{\vec}[1]{\mathbf{#1}}
\usepackage{tikz}

\begin{document}

\title{Connecting single-layer $t$-$J$ to Kondo lattice models: Exploration with cold atoms}

\author{Hannah Lange}
\affiliation{Department of Physics and Arnold Sommerfeld Center for Theoretical Physics (ASC), Ludwig-Maximilians-Universit\"at M\"unchen, Theresienstr. 37, M\"unchen D-80333, Germany}
\affiliation{Max-Planck-Institute for Quantum Optics, Hans-Kopfermann-Str.1, Garching D-85748, Germany}
\affiliation{Munich Center for Quantum Science and Technology, Schellingstr. 4, Munich D-80799, Germany}

\author{Eugene Demler}
\affiliation{Institute for Theoretical Physics, ETH Zurich, 8093 Zürich, Switzerland}

\author{Jan von Delft}
\affiliation{Department of Physics and Arnold Sommerfeld Center for Theoretical Physics (ASC), Ludwig-Maximilians-Universit\"at M\"unchen, Theresienstr. 37, M\"unchen D-80333, Germany}
\affiliation{Munich Center for Quantum Science and Technology, Schellingstr. 4, Munich D-80799, Germany}

\author{Annabelle Bohrdt}
\affiliation{Department of Physics and Arnold Sommerfeld Center for Theoretical Physics (ASC), Ludwig-Maximilians-Universit\"at M\"unchen, Theresienstr. 37, M\"unchen D-80333, Germany}
\affiliation{Munich Center for Quantum Science and Technology, Schellingstr. 4, Munich D-80799, Germany}

\author{Fabian Grusdt}
\affiliation{Department of Physics and Arnold Sommerfeld Center for Theoretical Physics (ASC), Ludwig-Maximilians-Universit\"at M\"unchen, Theresienstr. 37, M\"unchen D-80333, Germany}
\affiliation{Munich Center for Quantum Science and Technology, Schellingstr. 4, Munich D-80799, Germany}

\date{\today}
\begin{abstract}
The Kondo effect, a hallmark of many-body physics, emerges from the antiferromagnetic coupling between localized spins and conduction fermions, leading to a correlated many-body singlet state. Here we propose to use the mixed-dimensional ($\mathrm{mixD}$) bilayer Hubbard geometry as a platform to study Kondo lattice physics with current ultracold atom experiments. At experimentally feasible temperatures, we predict that key features of the Kondo effect can be observed, including formation of the Kondo cloud around a single impurity and the competition of singlet formation with Ruderman-Kittel-Kasuya-Yosida (RKKY) interactions for multiple impurities, summarized in the Doniach phase diagram. Moreover, we show that the mixD platform provides a natural bridge between the Doniach phase diagram of the Kondo lattice model, relevant to heavy-fermion materials, and the phase diagram of cuprate superconductors as described by a single-layer Zhang-Rice type $t$-$J$ model: It is possible to continuously tune between the two regimes by changing the interlayer Kondo coupling. Our findings demonstrate that the direct connection between high-temperature superconductivity and heavy-fermion physics can be experimentally studied using currently available quantum simulation platforms.
\end{abstract}
\maketitle

The Kondo effect is a ubiquitous phenomenon in electron systems, distinguished by a complexity of emergent many-body correlations and the possibility of accurate theoretical analysis. 
The essence of the Kondo effect is that the antiferromagnetic (AFM) coupling between a localized impurity spin and itinerant
conduction fermions leads to the
formation of a many-body singlet state, accompanied by
non-trivial renormalization of system parameters and the
emergence of a characteristic energy scale known as the
Kondo temperature. Originally proposed in the context
of metals with dilute magnetic impurities~\cite{Anderson1961,Kondo1964}, the
Kondo effect was shown to play a key role in the physics of heavy fermion materials~\cite{Si2010review} and transport phenomena in mesoscopic systems \cite{Glazman1988}. Possible connections between the Kondo physics and properties of cuprate superconductors have also been discussed, see e.g. Ref.~\cite{Cooper2022pseudogap}.

\begin{figure}[t]
\centering
\includegraphics[width=0.49\textwidth]{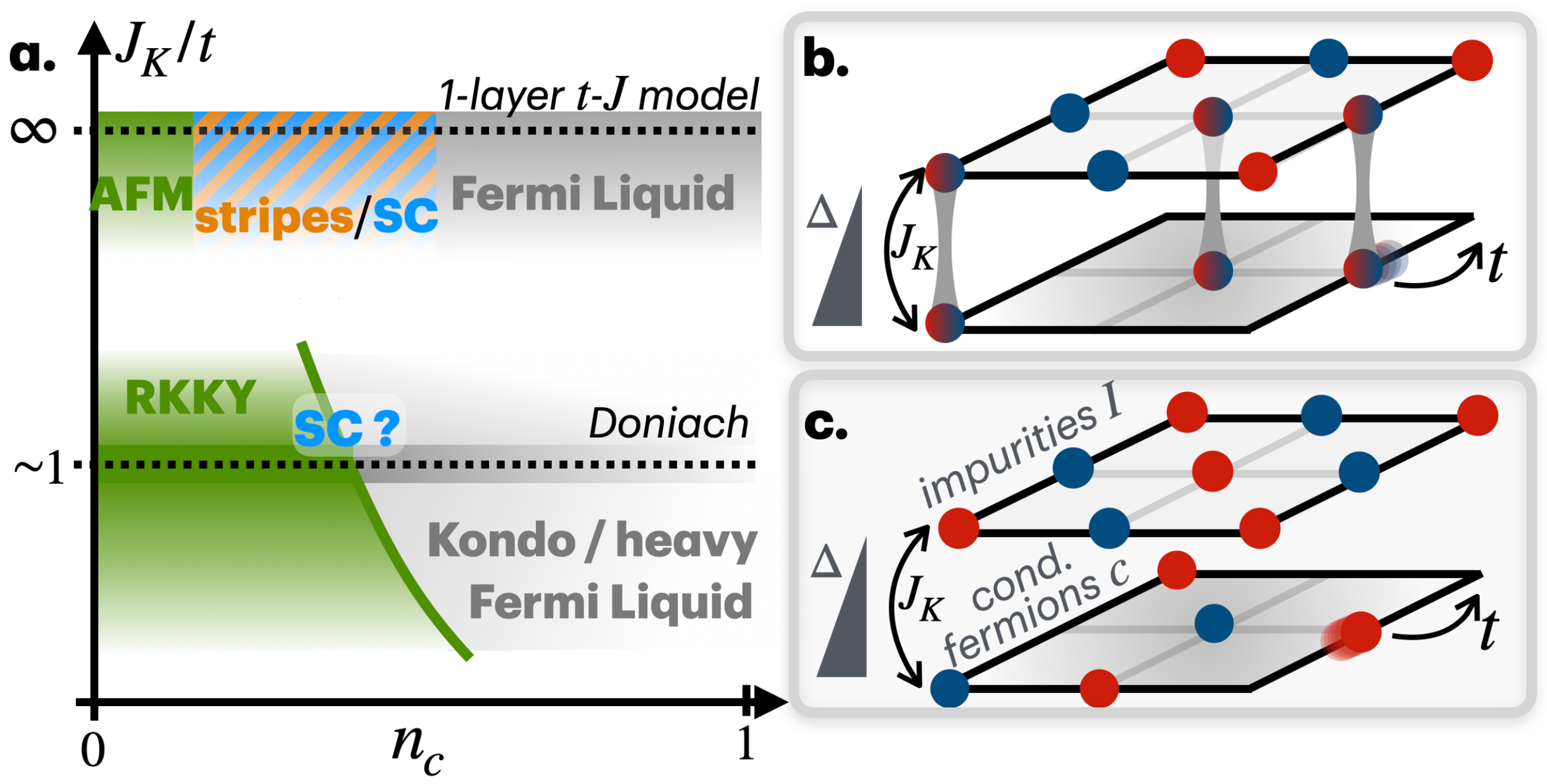}
\caption{ \textbf{a.} Exploration of Kondo lattice (KL) physics (bottom) and single-layer $t$-$J$ models (top), within a single mixD bilayer setup. By implementing a strong potential gradient $\Delta$ between the impurity (upper) and conduction (lower) layer, a widely tunable Kondo interaction $J_K(\Delta)$ can be realized for cold atoms. \textbf{b.} For strong $J_K\gg t$, singlets form on each rung and effectively behave as dopants in the magnetic background of a single-layer model. \textbf{c.} For smaller $J_K$, the KL with its underlying competition between Kondo and RKKY physics can be explored, including in the experimentally feasible regime with on-site Hubbard interactions in both layers.}
\vspace{-0.1cm}
\label{fig:overview}
\end{figure}

In recent years, ultracold atoms in optical lattices~\cite{Bloch2008,Gross2017} have become powerful and highly tunable platforms for simulating strongly correlated quantum systems. Several theoretical proposals have outlined the realization of Kondo models using alkaline-earth atoms, exploiting their electronic structure and tunable interactions to mimic the AFM coupling between a conduction bath and localized impurities~\cite{FossFeig2010,Nakagawa2015,Zhang2016,KanaszNagy2018}. Here, we propose an alternative route, one that is independent of atomic species and operates at energy scales accessible to current quantum simulators. Specifically, we consider a mixed-dimensional (mixD) bilayer or ladder model~\cite{Hirthe2023,Bohrdt2021,Bohrdt2022}, see Fig.~\ref{fig:overview}, as similarly implemented in recent cold atom experiments~\cite{Hirthe2023,bourgund2023formation}, and make use of its similarity to Kondo lattice model (KLM) and Kondo-Heisenberg models \cite{Coleman2015}. In most of the discussion, we focus on two-dimensional (2D) bilayers in order to emphasize the connection of our setup to the physics of layered materials such as cuprates. However, most of our arguments also apply to the case of ladders for which we present numerical results from matrix product states (MPS) in the second part of the paper, that support the arguments developed before.

As we propose in this work, the mixD setup paves the way for the experimental observation of hallmark features of Kondo physics in the strong-coupling regime, as well as the physics associated with cuprate compounds, all in one setup that is well within reach of modern quantum simulators. Thereby we provide a new perspective on the connection between the KLM and unconventional superconductivity, which has been of interest to numerous works including Refs. \cite{Sikkema1997,Ohkawa2004,Tsvelik2016,Chang_2025}. Our numerical and perturbative analysis predicts that, by tuning the coupling strength $J_K$ between the layers, a connection between single-layer models (related to cuprates) and the KLM (relevant to heavy-fermion materials) can be experimentally observed. Specifically, for strong inter-layer coupling $J_K \gg t$, the setup maps exactly onto the single-layer Zhang-Rice type $t$-$J$ model \cite{schloemer2024localcontrolmixeddimensions,Sikkema1997} which can be derived from the three-band model of Cu-O layers in cuprate superconductors~\cite{Zhang1998,Emery1988}, see Fig.~\ref{fig:overview}\textbf{b}. For smaller coupling $J_K \lesssim t$, we identify the competition between Kondo singlet formation of impurity spins with conduction fermions and Ruderman-Kittel-Kasuya-Yosida (RKKY) interactions~\cite{Ruderman1954,Kasuya1956,Yosida1957} mediated by the conduction band, see Fig.~\ref{fig:overview}\textbf{c}, giving rise to the Doniach-type phase diagram~\cite{Doniach1977}. How these paradigmatic limits are connected at intermediate $J_K$ can be explored in today's quantum simulation platforms. Taken together, this positions the mixD model as a unified and experimentally accessible framework for exploring a broad class of strongly correlated systems with ultracold atoms.\\ 

\textit{Model.---}
Our starting point is a single-layer Fermi-Hubbard model, 
\begin{align}
\mathcal{\hat{H}}^\mu_\mathrm{FH}(t_\mu,U_\mu, \Delta_\mu) =& -t_\mu \sum_{\langle \mathbf{i},\mathbf{j}\rangle} \left( \hat{c}_{\mu,\mathbf{i}}^\dagger \hat{c}_{\mu,\mathbf{j}} + \mathrm{h.c.} \right) \notag \\
&+ U_\mu \sum_{i} \hat{n}^\mu_{\mathbf{i}\uparrow} \hat{n}^\mu_{\mathbf{i}\downarrow}+\Delta_\mu \sum_{i} \hat{n}^\mu_{\mathbf{i}} ,
\label{eq:FH}
\end{align}
where $t_\mu$ is the hopping amplitude, $U_\mu$ the on-site interaction strength and $\Delta_\mu$ the chemical potential of layer $\mu$. This model has been realized in various quantum simulation platforms~\cite{Bloch2008,Gross2017}. Two such layers, $\mu=I,c$, are then coupled via interlayer hopping $t_{cI}$, as also already realized in cold atom platforms~\cite{Gall2021,Koepsel2020}. Throughout the paper, we set the $I$-layer to half-filling, $n_I=1$, and consider varying conduction filling $n_c$. We identify one layer as the conduction layer~$c$ and the other as impurity layer~$I$.

A potential offset $\Delta=\Delta_c-\Delta_I$ between the layers is introduced to obtain a mixD situation, see Fig.~\ref{fig:overview}: For $\vert U_c \pm \Delta\vert, \vert \Delta \vert \gg t_{cI}$, direct hopping between the layers is suppressed and virtual hopping generates an interlayer spin-exchange interaction of Kondo type, with strength~\cite{Trotzky2008,Bohrdt2021}:
\begin{align}
J_K(\Delta) = 2 t_{cI}^2 \left( \frac{1}{U_c + \Delta} + \frac{1}{U_I - \Delta} \right).
\label{eq:JK}
\end{align}
For $U_c, U_I\gg t_c,t_I$, double occupancies in both layers are suppressed, giving rise to a mixD $t$-$J$ model~\cite{Hirthe2023,Bohrdt2021,Bohrdt2022}:
\begin{align}
\mathcal{\hat{H}}_{\mathrm{mixD}} = \mathcal{\hat{H}}^c_{tJ} + \mathcal{\hat{H}}^I_{tJ} + J_K \sum_{\mathbf{i} \in \mathcal{N}_I}\left(\hat{\vec{S}}_\mathbf{i}^I \cdot  \hat{\vec{S}}_\mathbf{i}^c-\frac{1}{4} \hat{n}^I_{\vec{i}} \hat{n}^c_{\vec{i}} \right).
\label{eq:mixDtJ}
\end{align}
Here, each $t$-$J$ Hamiltonian is defined as
\begin{multline}
\mathcal{\hat{H}}^\mu_{tJ} (t_\mu, J_\mu) = -t_\mu \hat{\mathcal{P}} \sum_{\langle \mathbf{i},\mathbf{j}\rangle} \left( \hat{c}_{\mu,\mathbf{i}}^\dagger \hat{c}_{\mu,\mathbf{j}} + \mathrm{h.c.} \right) \hat{\mathcal{P}} \\
+ J_\mu \sum_{\langle \mathbf{i},\mathbf{j}\rangle} \left( \hat{\vec{S}}^\mu_\mathbf{i} \cdot \hat{\vec{S}}^\mu_\mathbf{j} - \frac{1}{4} \hat{n}^\mu_{\vec{i}} \hat{n}^\mu_{\vec{j}} \right) + \mathcal{\hat{H}}_{\rm 3s},
\label{eq:tJ}
\end{multline}
with $J_\mu = 4t_\mu^2/U_\mu$, the Gutzwiller projector $\hat{\mathcal{P}}$ enforcing maximum single occupancy and $\mathcal{\hat{H}}_{\rm 3s}$ a three-site term. Note that Eq. \eqref{eq:mixDtJ} is similar to the Kondo-Heisenberg model (KHM), which is studied, e.g., in the context of cuprate \cite{Andrei1989,Berg2010,Senthil2003,Tsvelik2016} and heavy-fermion superconductivity \cite{Paul2007,Liu_2014,Zhang2011,Bernhard2015}. For a single localized impurity, it is close to the interacting bath models, as e.g. considered for 1D systems in Refs. \cite{Furusaki1994,Costamagna2006}. In most of the following discussion in this first part of our paper, we will consider the case $U_c=U_I  =: U\gg t_I= t_c=: t$, which can be realized naturally in cold atom experiments. 
In the second part, when we focus on ladder systems, we will turn to $t_I=0$, closer to the canonical Kondo model, and explain how this limit can be readily realized in experiments.

\textit{Phase diagram.---} The ratio $J_K/t$ of the Kondo coupling in Eq.~\eqref{eq:JK} and the intralayer tunneling can be relatively freely tuned from values $J_K/t \ll 1$ to $J_K/t \gg 1$ in optical lattices. As we show now, this allows to continuously connect within one unified phase diagram the physics of the single-layer $t$-$J$ model, as in cuprates, to that of heavy fermions in a KLM, see Fig.~\ref{fig:overview}. We focus on $n_c=1$ and $n_I<1$.

In the strong-coupling regime \( J_K \gg t \), singlets form on every occupied rung, see Fig.~\ref{fig:overview}\textbf{b}. These singlets behave as mobile, fermionic dopants in the magnetic background of the upper $I$-layer, and can be viewed in analogy to the Zhang-Rice singlets in the three-band model of the cuprate layers \cite{Zhang1998,Emery1988}. The motion of the itinerant $c$-fermions is projected onto the rung-singlet subspace, which leads to an effective tunneling $t_{\rm eff}=t/2$ of the singlets that displaces $I$-fermions along the way. Hence the low-energy physics corresponds to a single-layer $t$-$J$ model, Eq.~\eqref{eq:tJ}, with tunneling $t_{\rm eff}$, exchange $J$ and a modified three-site term (see supplemental material (SM)~\cite{SM} and Ref.~\cite{schloemer2024localcontrolmixeddimensions}), where rung-singlets take the role of doped holes. The doping $\delta = n_c$ is controlled by the density of $c$-layer fermions. 
Compared to a conventional implementation of the single-layer $t$-$J$ model as the large-$U$ limit of a single-band Hubbard model, this mapping has the advantage that dopants correspond to actual $c$-particles, thus providing direct experimental access to the properties of individual charge carriers~\cite{schloemer2024localcontrolmixeddimensions}. An interesting variation of the above model for $t_c, t_I, U_I \ll U_c$ allows to separate experimentally the effects of doping and doublon-hole fluctuations, as we discuss further in the SM \cite{SM}.


Returning to the experimentally most natural case when $U_c=U_I=U$ and $t_c=t_I=t$, we discuss what happens when $J_K$ is decreased, moving downward in the phase diagram in Fig.~\ref{fig:overview}\textbf{a}. Here, the probability of rung singlet formation decreases and a reduction to an effective single-layer model is no longer possible. Still assuming $U \gg t$ but for $J_K < t$ and for low conduction filling, the system reduces to a model that is very close to the canonical KLM: the exchange interaction between the local moments forming in the $I$-layer becomes vanishingly small, $J\to0$, when $U/t \to \infty$ and, although the $c$-fermions remain strongly interacting, they can form a Fermi liquid (at $J_K=0$) when their density $n_c$ is sufficiently small (e.g. $n_c \lesssim 0.7$ for $U/t \approx 8$~\cite{Koepsell2021}). As described in the Doniach phase diagram, at $0<J_K<t$ this state can either evolve into a heavy Fermi liquid, or form a magnetically ordered state, see Fig.~\ref{fig:overview}\textbf{a,c}. 

For the $n_c<1$ that we consider and small $J_K$, conduction fermions mediate effective Heisenberg interactions between the local moments formed in the $I$-layer. The resulting long-range coupling at distance $r$ is given by 
\begin{align}
J_\mathrm{RKKY}(r) \propto J_K^2 \rho_c \frac{\cos(2k_F r)}{|r|^d},
\label{eq:RKKY}
\end{align}
with $d$ the dimension of the system, $k_F$ the Fermi momentum and $\rho_c$ the density of states of the $c$-fermions at the Fermi energy. This yields a characteristic energy scale $E_\mathrm{RKKY} \propto J_K^2 \rho_c$~\cite{Coleman2015} of RKKY interactions, which compete with Kondo singlet formation. The characteristic energy scale associated with the latter -- the Kondo temperature $T_K$ -- can be derived for a single impurity via a scaling approach,
\begin{align}
    T_K \propto D_c \exp\left(-\frac{1}{\alpha \rho_c J_K}\right),
    \label{eq:TK}
\end{align}
where $\alpha = 1$ and $D_c\propto t_c$ is the conduction band width~\cite{Coleman2015}. For multiple impurities forming a regular lattice -- when the KLM is realized -- the effective energy scale is enhanced, characterized by $\alpha > 1$~\cite{Rice1986,Tsunetsugu1992}. The competition between $T_K$ and $E_\mathrm{RKKY}$ leads to the Doniach phase diagram, with a magnetically ordered state dominated by RKKY interactions at small $\rho_c J_K$ and a heavy Fermi liquid including Kondo singlets when $\rho_c J_K$ is large. Since $\rho_c$ increases monotonically with $n_c$, even at large $U_c/t_c \gg 1$ when $n_c \lesssim 0.8$~\cite{bulut1996lowenergyelectronicexcitationslayered,Khatami2011,Sordi2012}, we expect a transition from AFM to heavy Fermi liquid as $n_c$ is increased in the lower part of Fig.~\ref{fig:overview}\textbf{a}. The same is true when $J_K$ is increased at fixed $n_c$.

To conclude the first part of this letter, we proposed a new experimental platform that can establish a link between the phase diagrams of the KLM and the cuprate superconductors, both of which can be explored within the same, experimentally accessible mixD framework introduced in this work: In particular, both systems involve a transition from an AFM to a (heavy) Fermi liquid upon doping $\delta=n_c$. We conjecture that the transition  is governed by the same universal (Kondo-destruction) quantum critical point -- as supported e.g. by the common strange-metal behavior reported in both systems~\cite{Hu2024}. Exploring this connection in the mixD bilayer Hubbard model using ultracold atoms constitutes a formidable task for analog quantum simulators. \\

\textit{Numerical results for ladders.---}
In the following, we turn to the 1D setting of mixD ladders in order to facilitate numerical simulations with MPS. Specifically, we time-evolve MPS~\cite{PAECKEL2019167998} using the two-site version of the time-dependent variational principle (TDVP) implemented in SyTen~\cite{syten1,syten2}, see SM~\cite{SM}, to access the dynamics and finite temperature behavior. We demonstrate that the distinct regimes of the phase diagram can be observed even in this minimal 1D setup: In particular, (i) Kondo singlet formation and (ii) its competition with RKKY physics can be readily explored in today's quantum simulation platforms. Although we focus on the experimentally most natural case when conduction fermions are strongly interacting, $U_c \gg t_c$, and Kondo couplings are sizable, $J_K \simeq t_c$ leading to a breakdown of scaling analysis, we find that the competition between Kondo singlet formation and RKKY interactions remains intact. Together with the recent achievement of ultra-low temperatures, down to a few percent of tunneling energy, in Hubbard simulations \cite{Xu2025}, this sets the stage for future experimental exploration of the entire phase diagram in Fig.~\ref{fig:overview}. The strong $J_K/t$ regime, where the single-layer cuprate physics emerges, has been numerically observed in Ref. \cite{schloemer2024localcontrolmixeddimensions} and will not be further studied here. 

\begin{figure}[t]
\centering
\includegraphics[width=0.49\textwidth]{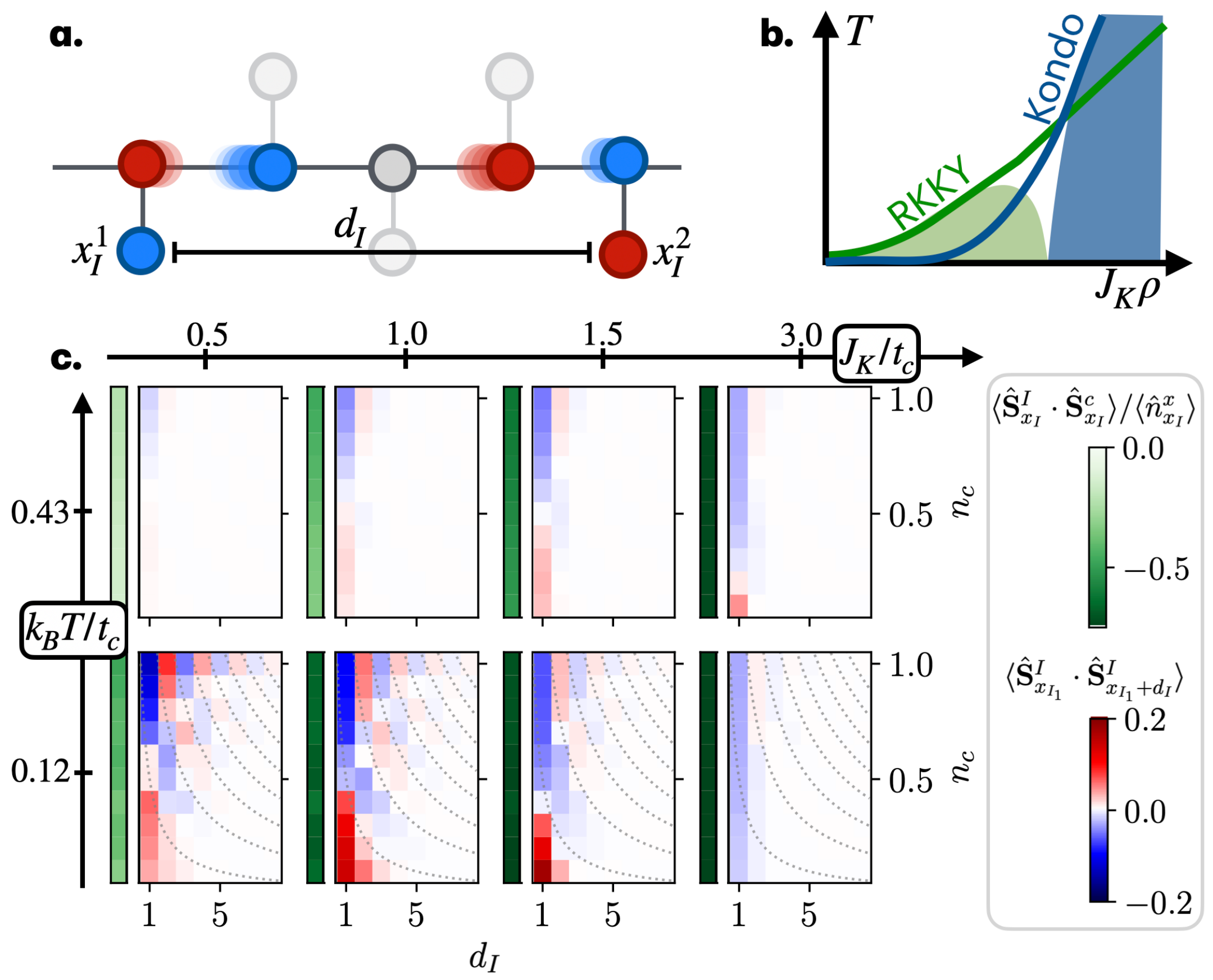}
\caption{RKKY vs. Kondo for two impurities coupled to a $t$-$J$ chain at finite temperatures $T$. \textbf{a}. The proposed setup, with impurity sites placed alternatingly below and above the conduction fermions to suppress direct couplings. \textbf{b}. Doniach phase diagram, with the RKKY (green) and Kondo (blue) regimes. \textbf{c}. Kondo signal $\langle \hat{\vec{S}}_{x_I}^I\cdot \hat{\vec{S}}_{x_I}^c\rangle $ (green) as well as the RKKY signal $\langle \hat{\vec{S}}_{x_{I_1}}^I\cdot \hat{\vec{S}}_{x_{I_2}}^I\rangle $ (blue and red), for different temperatures $k_BT$ and Kondo couplings $J_K$, with $J_c/t_c=0.5$, $J_I/t_c=0$ and length $L=20$. For the former, we average over two impurities with distance $d_I=13$; for the latter, we fix $x_{I_1}=4$ and consider different distances $d_I=x_{I_1}-x_{I_2}$. Dotted gray lines indicate $d_I^*$, where a sign change is expected from Eq.~\eqref{eq:RKKY} for a non-interacting bath. }
\label{fig:2imps}
\end{figure}

In 1D, the experimental setup described above can be adapted in order to achieve $t_I=0$ and $J_I=0$. This is possible by arranging the impurities alternatingly above and below the conduction chain (see Fig.~\ref{fig:2imps}a). Furthermore, digital micromirror devices (DMDs) can be used in order to shape the lattice potential such that direct couplings between impurity sites are fully suppressed.

In the remainder of the paper, we will first consider the dynamical singlet formation in the experimentally realizable setup, namely the 1D Fermi-Hubbard system with potential offset $\Delta$ between the bath and impurities, before turning to the finite-temperature equilibrium Kondo signatures beyond single impurities in the effective 1D mixD $t$-$J$ model. 

\begin{figure}[b]
\centering
\includegraphics[width=0.49\textwidth]{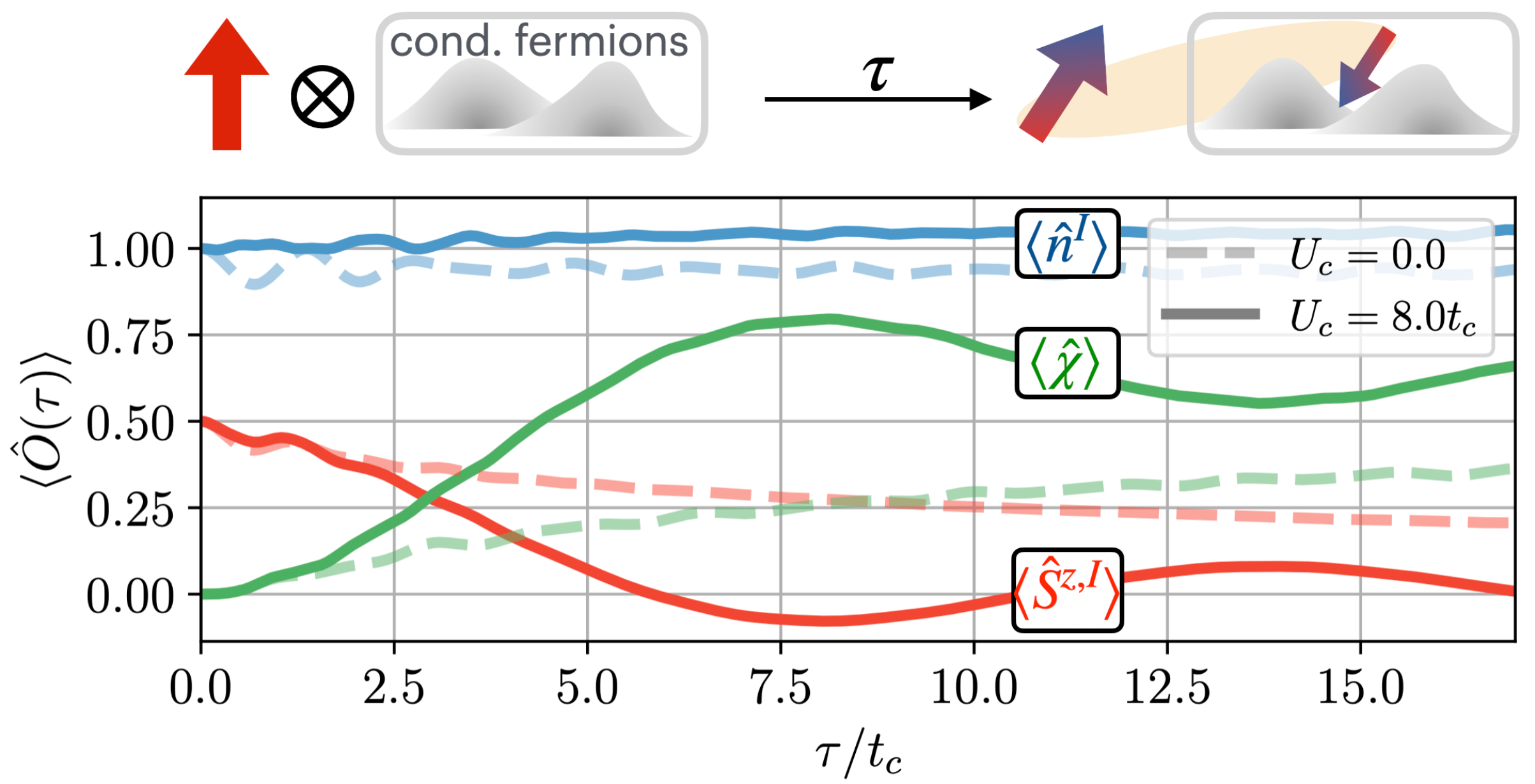}
\caption{Time evolution of the impurity density $\langle \hat{n}^I\rangle$, spin $\langle \hat{S}^{zI}\rangle$ and total spin-bath correlator $\langle \hat{\chi}\rangle$ after coupling to a bath of $L=20$ sites at time $\tau=0$ (see inset). Conduction baths with $U_c=U_I=8.0t_c$ (solid lines) and $U_c=0$ (dashed lines) are considered; in both cases $U_I=8.0t_c$, $\Delta=U_I/2$ and $n_c=0.7$. All calculations are for temperature $T=0$.}
\label{fig:FH}
\end{figure}

\textit{A single impurity in the microscopic model.--} 
We first present the simplest possible implementation that allows to study the dynamical formation of a Kondo singlet by monitoring the decay of the impurity spin: a one-dimensional (1D) conduction chain coupled to a single impurity located at site \( I = 0 \), using the full microscopic Hubbard model \eqref{eq:FH}. Specifically, we consider a conduction bath of length \( L = 20 \) at filling \( n_c = 0.7 \), with tunneling amplitude \( t = t_c = t_{cI} \), on-site interactions \( U_c = U_I = 8.0t \) and a potential offset $\Delta=U/2$, see Fig.~\ref{fig:FH}. We prepare the ground state of the chain and couple a single spin-up impurity to the chain at time $\tau=0$. An alternative experimental setting, which does not require the preparation of a single spin-polarized impurity, is presented in the Supplemental material (SM)~\cite{SM}.

Fig.~\ref{fig:FH} shows that, due to the large potential offset $\Delta=U/2\gg t_{cI}$, the impurity density $n_I$ remains almost constant at $n_I\approx 1$. The spin exchange interaction Eq.~\eqref{eq:JK} couples the impurity spin to the conduction fermions, as can be observed in a decrease of the local spin $\langle \hat{S}^{z,I}_{x_I}\rangle $~\cite{Bragancca2021,Wauters2024}. For comparison, we also show $U_c=0$ (canonical Kondo impurity problem), with an approximately monotonic decrease. For $U_c=8.0$ we observe a similar decrease in $\hat{S}_I^z$, but with oscillations, reflecting the enhanced AFM correlations in the bath in this case. 
Furthermore, we introduce
\begin{align}
    \hat{\chi}=-\sum_{x_c}\hat{\vec{S}}_{x_I}^I\cdot \hat{\vec{S}}^c_{x_c}
    \label{eq:chi}
\end{align}
to study how spin-correlations of the impurity with the bath emerge. Fig.~\ref{fig:FH} shows that $ \langle\hat{\chi}\rangle $ builds up with time $\tau$, providing a direct signature of Kondo singlet formation, accessible in experiments; $\langle \hat{\chi}\rangle =0.75$ would correspond to the development of a full spin singlet.

To conclude, non-equilibrium Kondo signatures are readily accessible in cold atom simulators and emerge for $U_c\gg t_c$. In the SM \cite{SM}, we further demonstrate that the Kondo cloud and characteristic spectral features persist under such strong interactions. This motivates us to study the corresponding effective model -- namely, the mixD Hamiltonian Eq.~\eqref{eq:mixDtJ} with $J_c/t_c=0.5$  and two impurities in the following. 

\textit{Two impurities in the effective model.---} For multiple impurities, the competition between RKKY and Kondo physics, schematically shown in the Doniach phase diagram in Fig.~\ref{fig:2imps}b, can be observed. In Fig.~\ref{fig:2imps}a, we consider the minimal mixD setup to observe this effect: two impurities coupled to a $t$-$J$ chain at distance $d_I$, with $J_I=0$, different conduction fermion fillings $n_c$ and Kondo interactions $J_K$ at temperatures $k_BT/t_c=0.12, 0.43$.

We focus on $k_BT/t_c=0.12$ first, which is on the order of temperatures that can be achieved with state-of-the-art cooling protocols in cold atom simulators \cite{Xu2025}: The Kondo regime is probed by $\langle \hat{\vec{S}}_{x_I}^I\cdot\hat{\vec{S}}_{x_I}^c \rangle$ on the same rung $x_I$ (green color scale in Fig.~\ref{fig:2imps}c). As expected, the signal becomes stronger with increasing $J_K/t_c$. The RKKY interactions are probed by spin correlations between the impurities, $\langle \hat{\vec{S}}_{x_{I_1}}^I\cdot \hat{\vec{S}}_{x_{I_2}}^I\rangle$, with different distances $d_I=x_{I_1}-x_{I_2}$ between the latter (blue / red color scale in Fig.~\ref{fig:2imps}c). In contrast to the Kondo signal, this quantity is strongest for the smaller $J_K \lesssim t_c$ and the sign of the signal depends on $d_I$ as well as the filling $n_c$. This is expected from Eq.~\eqref{eq:RKKY}, where the sign of $J_\mathrm{RKKY}$ depends on the distance as well as the Fermi momentum. For free fermions, $k_F=\pi n_c/2$, this implies that sign changes occur for $d_I^*=1/n_c(\mathbbm{Z}+1/2)$, which are indicated by gray lines in Fig.~\ref{fig:2imps}c. Although our conduction fermions are not free ($U_c/t_c=8$ for the considered $J_c/t_c=0.5$), the agreement is very good. In particular, this captures the sign change of $\langle \hat{\vec{S}}_{x_{I_1}}^I\cdot \hat{\vec{S}}_{x_{I_1}+1}^I\rangle$ (i.e. $d_I=1$) near $n_c\approx 0.5$, which is also observed for $J_K\lesssim  t_c$. For larger $J_K/t_c>1$, the sign change at $d_I=1$ shifts to smaller $n_c$, and is accompanied by a stronger Kondo singlet signal, potentially indicating that the RKKY regime ends. For the higher temperature $k_BT/t_c=0.43$, both Kondo and RKKY signatures remain observable, with weaker signals especially for $d_I>1$, and the strongest signal shifted to higher $J_K/t_c$. This is in agreement with the Doniach phase diagram Fig.~\ref{fig:2imps}b, where the onset of the RKKY regime is shifted to larger $J_K/t_c$ when the temperature is increased.\\


\textit{Conclusion and outlook.---} To conclude, the mixD $t$-$J$ bilayer provides a highly tunable platform for realizing single-layer physics relevant to cuprates as well as key features of the Kondo (lattice) model. We show that the two regimes and their connection can be readily accessed using cold atoms in optical lattices, simply by tuning $J_K$ using a potential offset between the layers. For the Kondo and RKKY regimes, our simulations reveal that hallmark features -- such as singlet formation, impurity screening, and their competition with RKKY interactions -- are accessible within experimentally achievable temperature regimes \cite{Xu2025}. 

These results pave the way for several promising research directions: While our numerical analysis reveals key features of Kondo and RKKY physics in 1D settings but is not easily extendable to 2D settings, the underlying experimental framework is directly applicable to 2D and large system sizes. This would allow to study the question whether high-$T_c$ $d$-wave superconductivity for single-layer models can be directly related to heavy-fermion superconductors located around the Kondo critical point, similar to observations for the Kondo-Heisenberg model \cite{Tsvelik2016,Paul2007}. Additionally, the Kondo effect could be explored via transport measurements in systems with randomly distributed impurities. Further extensions include the consideration of multiple conduction channels or higher spin quantum numbers. In particular, spin-1 Kondo physics becomes accessible by realizing $J_K<0$, which is straightforward in our setup.

Lastly, we would like to mention that the physics of bilayer nickelate superconductors can be studied in our mixD bilayer setup: For bilayer nickelate superconductors~\cite{Sun2023,zhang2023hightemperature}, the relation to the mixD model has been established in a range of works including Refs.~\cite{schloemer2023superconductivity,lange2023feshbach,lange2023pairing,borchia2025,Qu2023,Lu2023}. 

Our approach hence establishes a direct connection between the phase diagrams of Kondo systems and those of single-layer and multi-layer unconventional superconductors within a unified cold-atom setup. This platform thus offers a versatile setting for investigating a wide range of strongly correlated phenomena using ultracold atom quantum simulators.\\

\emph{Acknowledgements.---} We would like to thank Tizian Blatz, Immanuel Bloch, Markus Greiner, Chen-How Huang, Adam Kaufman, Philipp Preiss, Henning Schlömer, Liyang Qiu and Yi-Fan Qu for helpful discussions. We acknowledge funding by the Deutsche Forschungsgemeinschaft (DFG, German Research Foundation) under Germany's Excellence Strategy -- EXC-2111 -- 390814868 and from the European Research Council (ERC) under the European Union’s Horizon 2020 research and innovation programm (Grant Agreement no 948141) — ERC Starting Grant SimUcQuam. HL acknowledges support by the International Max Planck Research School for Quantum Science and Technology (IMPRS-QST). ED acknolwedges support from the SNSF project 200021\_212899, the Swiss State Secretariat for Education, Research and Innovation (contract number UeM019-1), and the ARO grant number W911NF-20-1-0163. Numerical simulations were performed on the Arnold Sommerfeld Center for Theoretical Physics High-Performance Computing cluster. \\

\bibliography{main}
\FloatBarrier
\newpage
\onecolumngrid
\widetext
\newpage
\appendix

\setcounter{equation}{0}
\setcounter{figure}{0}
\setcounter{table}{0}
\setcounter{page}{1}
\makeatletter
\renewcommand{\theequation}{S\arabic{equation}}
\renewcommand{\thefigure}{S\arabic{figure}}

\begin{center}
    \vspace{2em}
    {\large\bfseries End Matter \par}
    \vspace{1em}
\end{center}

\section{Cold atom realization}
\begin{figure}[htp]
\centering
\includegraphics[width=0.5\textwidth]{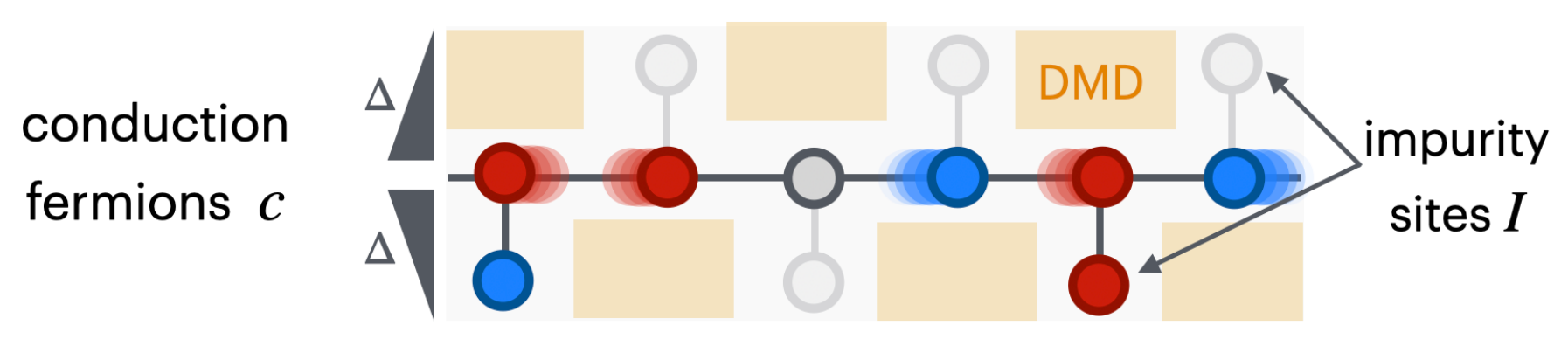}
\caption{The mixed-dimensional (mixD) bilayer setup: The mixD setup realizing a Kondo model in 1D with one or multiple impurities, arranged alternatingly below and above
the conduction chain to suppress direct nearest-neighbor impurity coupling. By implementing a strong gradient $\Delta$ between impurity and conduction sites a Kondo coupling with suppressed tunneling can be realized (for potential $\Delta>0$ realizing the Anderson impurity model). The impurities can be localized by shaping the potential landscape with a digital mirror device (DMD, orange).}
\label{fig:setup}
\end{figure}

We are interested in the following class of models,
\begin{align}\mathcal{\hat{H}}_\mathrm{K}= \mathcal{\hat{H}}^c_\mathrm{FH}(t_c,U_c) + \mathcal{\hat{H}}^I_\mathrm{FH}(t_I, U_I) + J_K \sum_{\mathbf{i}\in \mathcal{N}_I} \hat{\vec{S}}_\mathbf{i}^I \cdot \hat{\vec{S}}_\mathbf{i}^c,\label{eq:Kondo}\end{align} with the Fermi-Hubbard Hamiltonian $\mathcal{\hat{H}}_\mathrm{FH}$.
As in the main text, we identify one of the layers with impurities, on sites $\mathcal{N}_I=\{\mathbf{j}\,\vert \,\mathrm{impurity}\,\mathrm{at}\, \mathbf{j}\}$, and one with the conduction fermions, labeled by the index $\mu=I$ and $\mu=c$ respectively. The corresponding densities are denoted by $n_\mu=\frac{1}{N_\mu}\sum_{i=1}^{N_\mu}\langle \hat{n}_{i\uparrow}^\mu+\hat{n}_{i\downarrow}^\mu\rangle$, where $N_\mu=\vert \mathcal{N}_\mu\vert$ is the number of sites of type $\mu$. For $n_I = 1$, we vary the number of impurities by varying the number of impurity sites $N_{I}$, with particular emphasis on a single ($N_I=1$) and two impurities ($N_I=2$).

Our starting point for the experiment is a single-layer Fermi-Hubbard model, as realized in various quantum simulation platforms~\cite{Bloch2008,Gross2017}. Two such layers are then coupled via interlayer hopping $t_{cI}$. Bilayer systems of this kind have already been experimentally realized in cold atom platforms~\cite{Gall2021,Koepsel2020}. We refer to one layer as the conduction layer $c$ and the other as impurity layer $I$. A potential offset $\Delta$ between layers is introduced, see Fig.~\ref{fig:overview}a. For $\vert U_c + \Delta\vert , \vert U_I - \Delta \vert, \vert \Delta \vert \gg t_{cI}$, direct hopping between layers is suppressed, $t_{cI}^{\mathrm{eff}} \approx 0$. Virtual hopping processes generate an interlayer spin-exchange interaction of Kondo type, see Eq.~\eqref{eq:JK}~\cite{Hirthe2023,Bohrdt2021,Bohrdt2022}
reproducing the Kondo Hamiltonian in Eq.\eqref{eq:Kondo}. While for $\Delta>0 $ (and $U_c=0$), this realizes the Anderson impurity model~\cite{Anderson1961}, the setup also permits the exploration of metastable states with $\Delta<0$, which can be probed on finite time scales~\cite{Hirthe2023}. \\


To suppress direct impurity interactions, we propose two slightly different setups for the 1D and 2D cases: In 1D, we propose using digital micromirror devices (DMDs) to shape the optical lattice potential such that the impurity tunneling is suppressed to $t_I= 0$. Arranging the impurities alternatingly above and below the conduction layer (see Fig.~\ref{fig:setup}) localizes the impurities. This configuration can be readily implemented in current 1D quantum simulators. In the 2D bilayer setting, where it is much harder to shape the impurity potential by a DMD, we will consider a regime where $t_I$ leads to nearest-neighbor $J_I=4t_I^2/U$ between the impurities, see Eq.~\eqref{eq:tJ}, in addition to the RKKY interactions \cite{Zhang2011,Coqblin2003,Bernhard2015}. By realizing optical lattices with independently tunable parameters $t_\mu,U_\mu$ in the two layers, $J_I$ and $J_{\mathrm{RKKY}}$ could be tuned independently. 

\section{Alternative experimental procedure for the Kondo singlet formation in Fig.~\ref{fig:FH} \label{appendix:Experiment2} }
\begin{figure}[htp]
\centering
\includegraphics[width=0.8\textwidth]{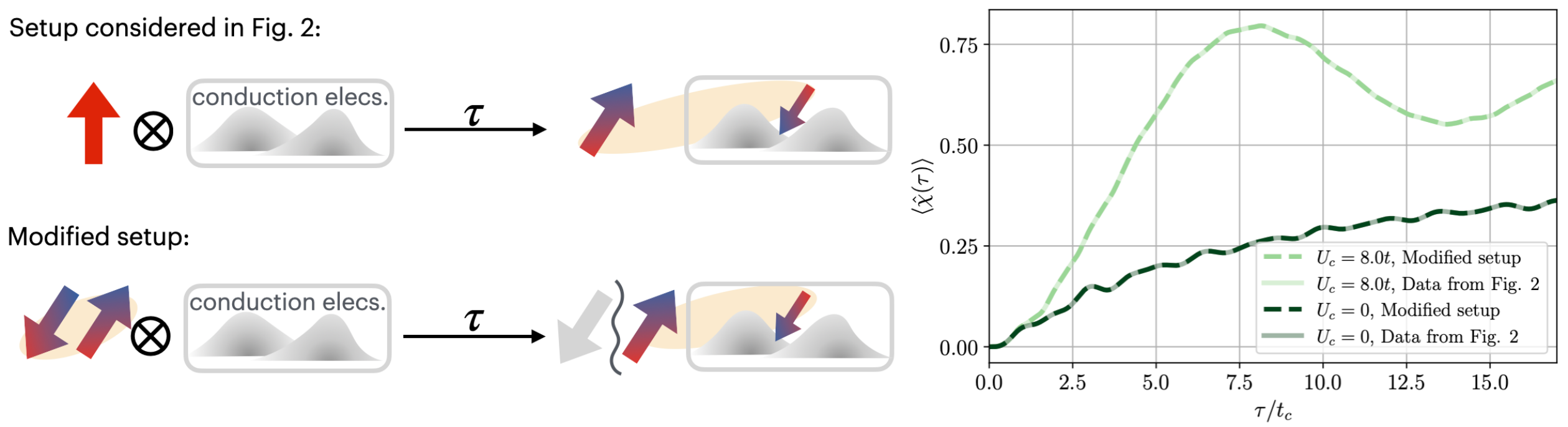}
\caption{Dynamical singlet formation experiment: \textbf{a}. Two experimental realizations are possible: (i) The one proposed in the main text and Fig.~\ref{fig:FH} (top), where a single spin up on the impurity site is prepared and then coupled to the bath at $\tau=0$. (ii) Alternatively, a singlet can be prepared and only one site is coupled to the bath (bottom). \textbf{b}. The dynamical singlet formation in both scenarios (light and dark green) leads to the same total spin-bath correlator $\langle \hat{\chi}(\tau)\rangle$, with the lines lying on top of each other.  Conduction baths with $U_c=U_I=8.0t_c$ (light green) and $U_c=0$ (dark green) are considered; in both cases $U_I=8.0t_c$, $\Delta=U_I/2$ and $n_c=0.7$. All calculations are for temperature $T=0$.}
\label{fig:FH2}
\end{figure}

For the dynamical singlet formation experiment in the main text and Fig.~\ref{fig:FH} an alternative experimental setup can be considered, that does not require the preparation of a spin-up at the impurity site, see Fig.~\ref{fig:FH2}a. In contrast to the setup proposed in the main text, a singlet can be prepared and only one site is then coupled to the bath as schematically shown in Fig.~\ref{fig:FH2}a (bottom). Fig.~\ref{fig:FH2}b shows that the dynamical singlet formation in both scenarios leads to the same total spin-bath correlator $\langle \hat{\chi}(\tau)\rangle$ defined in Eq.~\eqref{eq:chi}. 

\section{A single impurity in the effective model}

\begin{figure*}[htp]
\centering
\includegraphics[width=0.9\textwidth]{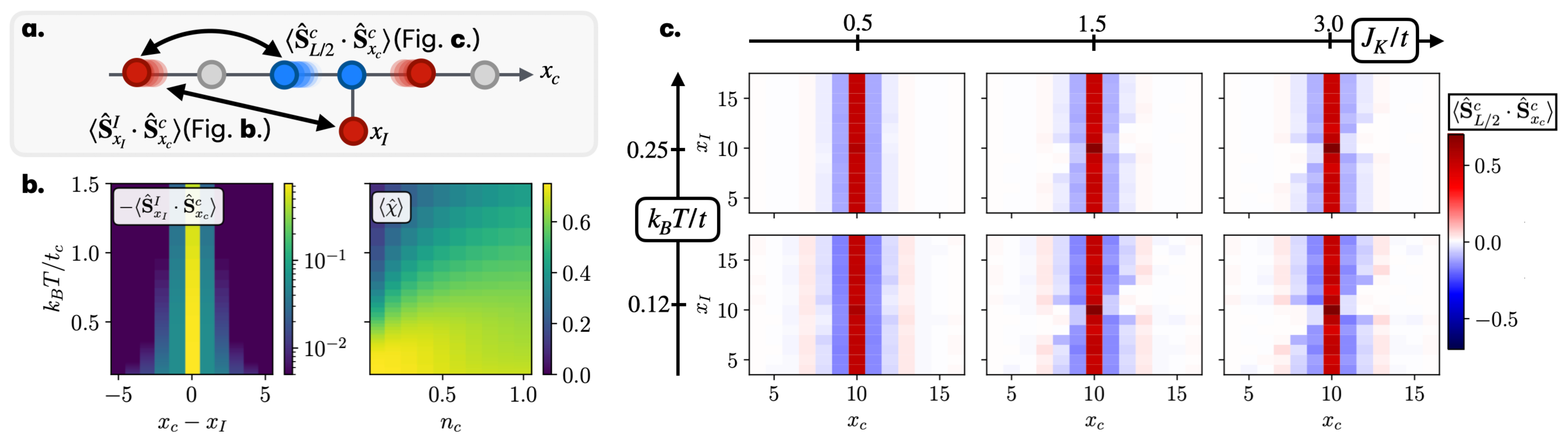}
\caption{Single Kondo impurity at site $x_I$, coupled to a $t$-$J$ chain with $J_c/t_c=0.5$ of length $L=21$ at filling $n_c=0.7$ and finite temperatures $T$. \textbf{a.} The setup. \textbf{b.} The spin-bath correlator $\langle \hat{\vec{S}}_{x_I}^I\cdot\hat{\vec{S}}_{x_c}^c \rangle$ for $n_c=1$ and $x_c=L/2$ (central site), and the summed signal $\langle \hat{\chi}\rangle$ from Eq.~\eqref{eq:chi} for different $n_c$, both for Kondo coupling $J_K/t_c=3.0$. \textbf{c.} Conduction correlations $\langle \hat{\vec{S}}_{L/2}^c\cdot\hat{\vec{S}}_{x_c}^c \rangle$ at $n_c=0.7$, which reveal the Kondo screening for all $x_c<x_I<L/2$ and $L/2<x_I<x_c$. The correlations considered in \textbf{b} and \textbf{c} are indicated in \textbf{a}.   }
\label{fig:1imp}
\end{figure*}

The mixD setup can be used to observe the Kondo cloud of a single impurity, along with the resulting screening effects. To demonstrate this, we present finite-temperature results for a single Kondo impurity coupled to a \( t \)-\( J \) chain of length \( L = 21 \), as illustrated in Fig.~\ref{fig:1imp}a. Similar to the previous section, we observe the formation of a Kondo singlet, and additionally show that the singlet character remains and can be observed also up to finite temperatures in a regime accessible by cold atom experiments~\cite{Xu2025}. In particular, in Fig.~\ref{fig:1imp}b (left) for an impurity at the central site $x_I=L/2$ and half-filled bath $n_c=1$ we observe a strong local singlet signal $\langle \hat{\vec{S}}_{x_I}^I\cdot\hat{\vec{S}}_{x_c}^c \rangle$ especially for $x_c=x_I$. The summed signal over all sites $\langle \hat{\chi}\rangle $ extends to a broader range of dopings; see Fig.~\ref{fig:1imp}b (right). 

The Kondo singlet formation also manifests through screening of interactions within the bath. This leads to a suppression of correlations $\langle \hat{\vec{S}}_{L/2}^c\cdot\hat{\vec{S}}_{x_c}^c \rangle$ between two conduction spins across the impurity at $x_I$, as demonstrated in Fig.~\ref{fig:1imp}c by varying $x_c$ and $x_I$; $L/2$ is the central site of the system. The $t$-$J$ chain without Kondo coupling features AFM correlations, which remain present up to the highest considered temperatures. Depending on the position of the impurity $x_I$ -- and the corresponding Kondo singlet -- the correlation $\langle \hat{\vec{S}}_{L/2}^c\cdot\hat{\vec{S}}_{x_c}^c \rangle$ can be strongly suppressed: In Fig.~\ref{fig:2imps}c, white triangular areas at low $T$ indicate the presence of a Kondo singlet and the corresponding screening of $\langle \hat{\vec{S}}_{L/2}^c\cdot\hat{\vec{S}}_{x_c}^c \rangle$. This can be understood as follows: Fixing the impurity to $x_I=L/2$, the conduction spin at $L/2$ becomes part of the Kondo singlet and all correlations of the spin at $x_I$ with spins at any $x_I\neq x_c$ are suppressed for large $J_K$. When shifting the impurity to the left $x_I<L/2$, all $\langle \hat{\vec{S}}_{L/2}^c\cdot\hat{\vec{S}}_{x_c}^c \rangle$  correlations for $x_c<x_I$ become vanishingly small, since the Kondo singlet sits at $x_I$ between the interacting conduction fermions at $L/2$ and $x_c$. In contrast, correlations for $x_I<x_c$ are not suppressed. The screening is clearly visible at $k_BT/t=0.12$, which is accessible with current quantum simulators~\cite{xu2025neutralatomhubbardquantumsimulator}, and signatures remain visible at $k_BT/t=0.25$ for $J_K>t$.

\newpage
\appendix
\begin{center}
    \vspace{2em}
    {\large\bfseries Supplemental Material \par}
    \vspace{2em}
\end{center}

\section{Additional results on the competition between RKKY and Kondo physics}
\begin{figure}[htp]
\centering
\includegraphics[width=0.9\textwidth]{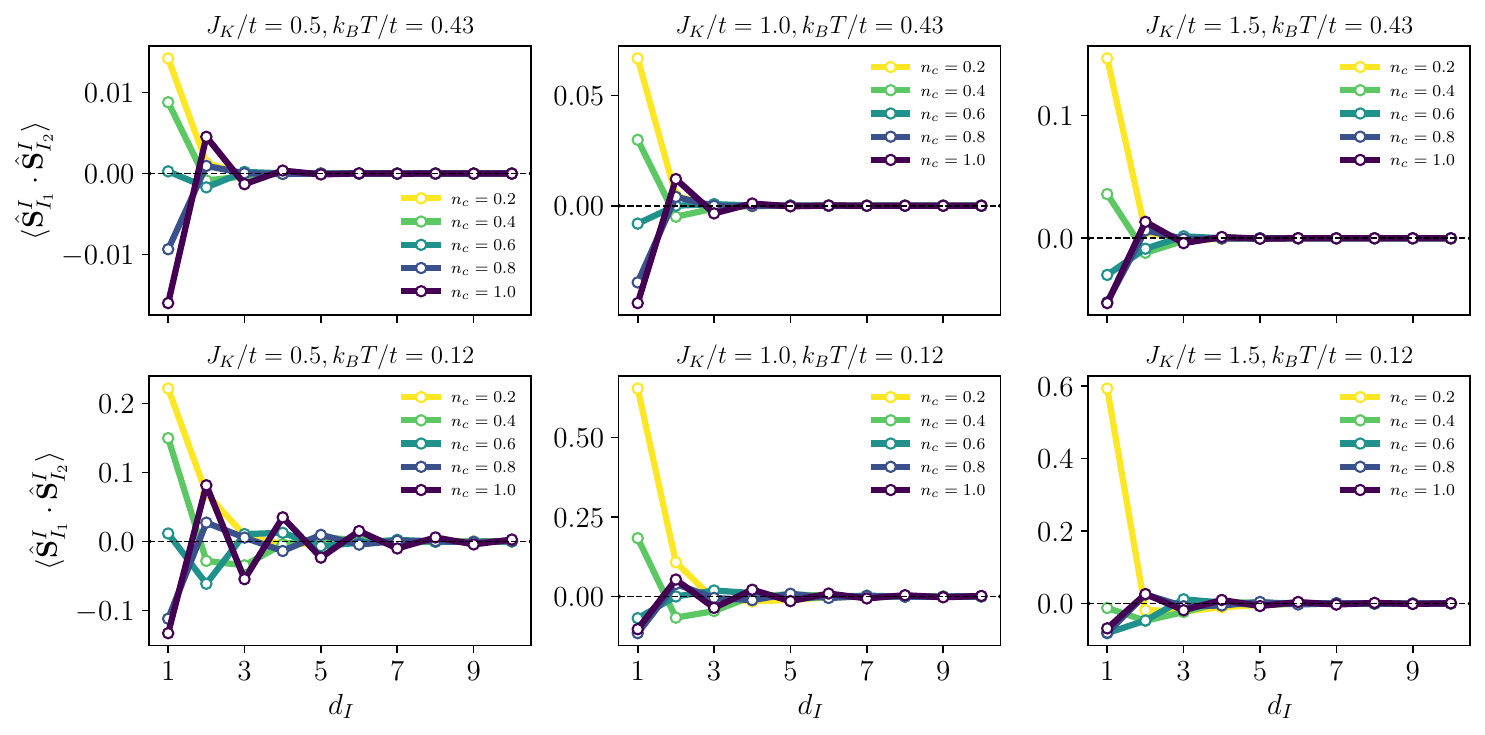}
\caption{ RKKY vs. Kondo: The RKKY signal $\langle \hat{\vec{S}}_{x_{I_1}}^I\cdot \hat{\vec{S}}_{x_{I_2}}^I\rangle $, with fixed $x_{I_1}=4$ and different distances $d_I=x_{I_1}-x_{I_2}$ between the impurities for different temperatures $k_BT$ (top vs. bottom) and Kondo couplings $J_K$ (left to right), with $J_c/t_c=0.5$ and length $L=20$.   }
\label{fig:2imps_cuts}
\end{figure}

For clarity, we show cuts through Fig. \ref{fig:2imps} for five exemplary fillings $n_c=0.2,0.4,0.6,0.8,1.0$ (indicated by the different colors) in Fig. \ref{fig:2imps_cuts}. As discussed in the main text, the sign of the signal depends both on the impurity distance $d_I$ and the filling $n_c$. For small $J_K\lesssim t$ the signal is clearly visible over large distances $d_I$.

\section{Kondo peak}
\begin{figure}[htp]
\centering
\includegraphics[width=0.5\textwidth]{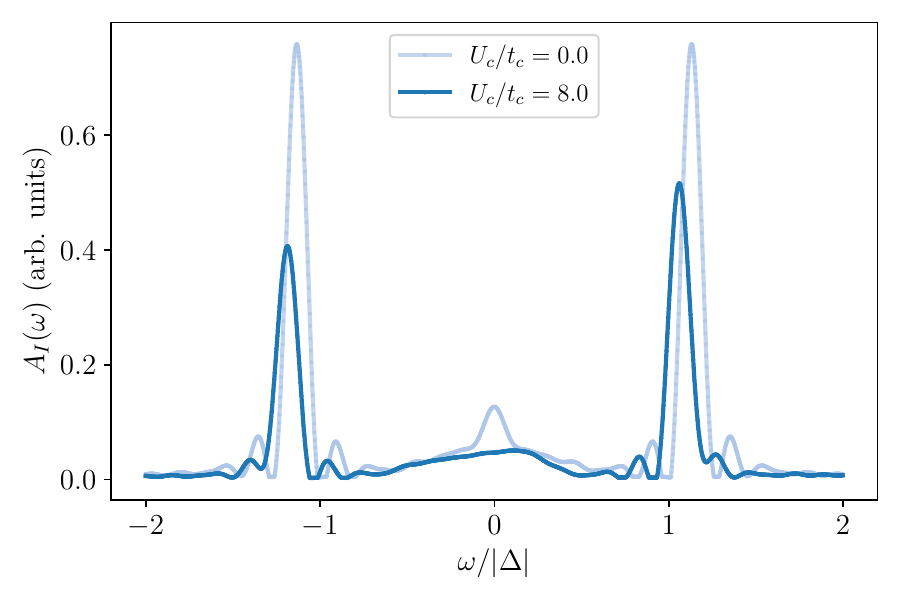}
\caption{The impurity spectral function $A_I(\omega)$ for a chain of $L=20$ sites coupled to a impurity at site $x_I=0$ with potential offset $\Delta=-U_I/2$ and $U=8.0t$. The contributions for $\omega<0$ ($\omega>0$) are dominated by the signal when the impurity is removed (an impurity is added at site $I$).   }
\label{fig:Aw}
\end{figure}

In Fig.~\ref{fig:Aw} we show the impurity spectral function $A_I(\omega)$ for a chain of length $L=20$ coupled to an impurity at site $I=0$. It is defined as the Fourier transform of
\begin{align}
    A_{I}(t) = \langle \psi \vert e^{i\mathcal{\hat{H}}t}\hat{c}^\dagger_{I\sigma}e^{-i\mathcal{\hat{H}}t}\hat{c}_{I\sigma}\vert \psi\rangle,
\end{align}
where $\hat{\mathcal{H}}$ is the Hamiltonian describing the full microscopic Fermi-Hubbard model with potential offset,  $\psi$ is the ground state, $\hat{c}_{I\sigma}^{(\dagger)}$ annihilate (create) a particle at the impurity site $I$. To calculate this quantity, the ground state $\psi$ has to be calculated, which can be done for $\Delta <0$ (in contrast to $\Delta >0$, where a metastable state is realized). Then, $\hat{c}_{I\sigma}^{(\dagger)}$ is applied and the resulting state is time evolved, see Sec.~\ref{appendix:timeevol}.\\

The impurity spectral function $A_I(\omega)$ exhibits the three characteristic peaks of the Kondo regime: two side peaks near $\omega = \Delta$ and $\omega = U+\Delta$ arising from valence fluctuations, and a central peak at $\omega=0$ associated with spin fluctuations. This central peak becomes significantly broader when interactions are introduced in the bath (see dark blue lines).\\ 

Spectral functions have been measured in fermionic quantum gas experiments~\cite{Jordens2008,Veeravalli2008,Chin2004,Stewart2008}. Since the measurement of $A_I(\omega)$ does not require any momentum resolution, a measurement scheme similar to the scanning tunneling microscopy protocol of Ref.~\cite{Kollath2007} could be applied.

\section{Simulation details}
\subsection{Time evolution \label{appendix:timeevol}}
We use a time evolution based on the two-site variant of the time-dependent variational principle (TDVP)~\cite{PAECKEL2019167998} implemented in Syten~\cite{syten1,syten2}. We use a time step of $\delta \tau =0.04 t_c$ up to $\tau = 30 t_c$. We use particle and total spin conservation and  consider bond dimensions up to $\chi=2048$.

\subsection{Finite temperature simulations \label{appendix:finiteT}}
For the finite temperature calculations we use purification schemes, and as for the ground state simulations all $U(1)$ symmetries are explicitly conserved, similar to Ref.~\cite{Schloemer2022}. Purification schemes are based on enlarging the Hilbert space by one auxiliary (ancilla) site per physical site, i.e. $N=N_a$, which allows to represent mixed states in the physical subset of the Hilbert space as pure states on the enlarged space. 

The finite temperature state is 
\begin{align}
    \ket{\Psi(\beta)}=\mathrm{exp}(-\beta \mathcal{H}/2)\ket{\Psi(\beta=0)},
\end{align}
with the maximally entangled state 
\begin{align}
    \ket{\Psi(\beta=0)} \propto \hat{\mathcal{P}}_{N=N_a}\left[ \otimes_{i=0}^{L_xL_y-1}\left( \ket{0,0}+\sum_\sigma \ket{\sigma \Bar{\sigma}}\right)\right],
\end{align}
where $\hat{\mathcal{P}}_{N=N_a}$ projects onto the subspace with $N=N_a$ and the first (second) entry in $\ket{\dots,\dots}$ refers to the physical (ancilla) state. Here, we take $\hat{\mathcal{H}}$ to be the mixD Hamiltonian \eqref{eq:mixDtJ}.

In practice, we have to find a MPS representation of the latter state and evolve it in imaginary time $\tau = \beta/2$~\cite{PAECKEL2019167998}. The former is done by a ground state search of a specifically tailored entangler Hamiltonian. We follow Ref.~\cite{Nocera2016} and use the following entangler Hamiltonians:
\begin{itemize}
    \item For the conduction fermion layer, we use the canonical $t$-$J$ entangler,
\begin{align}
    \hat{\mathcal{H}}^{t\mathrm{-}J}_\mathrm{ent}=-\sum_{i\neq j} \hat{\Delta}_i^\dagger \hat{\Delta}_j+\mathrm{h.c.},
\end{align}
with $\hat{\Delta}^\dagger_i = \frac{1}{\sqrt{2}}\left( \hat{c}_{i\uparrow} \hat{c}_{a(i)\downarrow}-\hat{c}_{i\downarrow} \hat{c}_{a(i)\uparrow}\right)$.
\item For the impurity layer and at half-filling of the conduction layer we use the grand-canonical Heisenberg spin entangler 
\begin{align}
    \hat{\mathcal{H}}^{\mathrm{Heis}}_\mathrm{ent}=-\sum_{i} \hat{S}_i^+ \hat{S}_{a(i)}^-+\mathrm{h.c.}\,.
\end{align}
\end{itemize}

Our calculations are done for system sizes $L_x=20$ and $\tau=4t_c$, i.e. the lowest temperature state after the time evolution has $k_BT/t_c=0.125$. We use particle and total spin conservation and  consider bond dimensions up to $\chi=3000$.\\

\begin{figure}[t]
\centering
\includegraphics[width=1.0\textwidth]{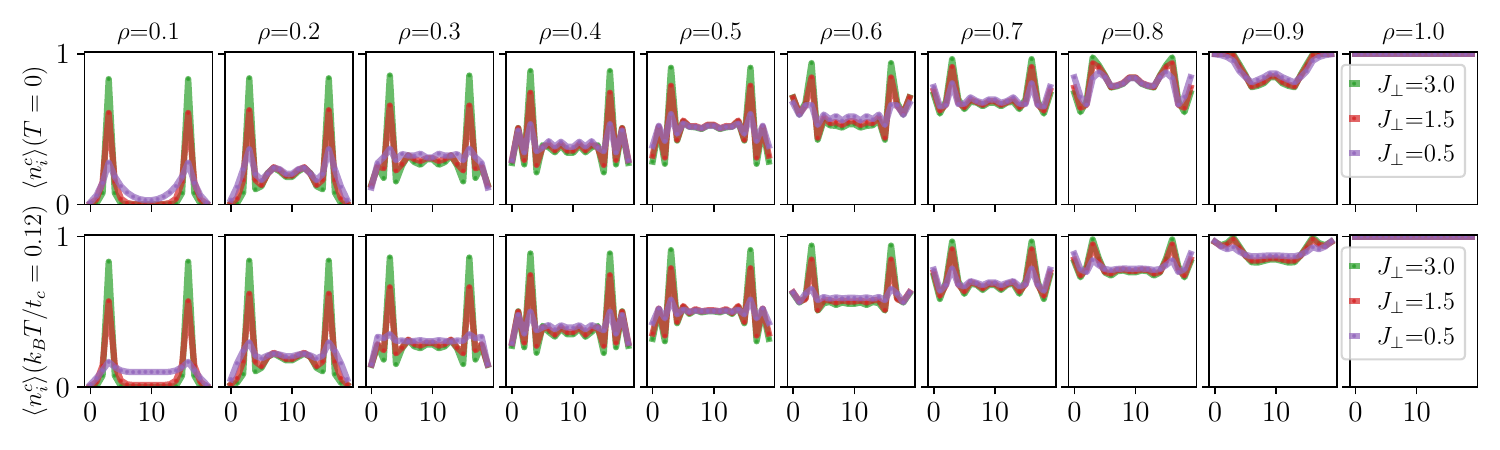}
\caption{Local density of ground state (top) and at $k_BT/t_c=0.125$ (bottom) for different conduction fillings $\rho$ and Kondo couplings $J_\perp$. We set $J_I=0$ and $J_c/t_c=0.5$.}
\label{fig:density}
\end{figure}

As shown in Fig.~\ref{fig:density}, the densities of lowest temperature state and ground state are very similar.

\section{Mapping to single-layer models}
Here, we show how the mixD bilayer model in the presence of dominant $J_K$ can be mapped to effective single-layer descriptions.
\subsection{Strong interactions $U_{c,I}\gg t_{c,I}$: Mapping to the single-layer Zhang-Rice type $t$-$J$ model}
As shown in Ref.~\cite{schloemer2023superconductivity}, the mixD bilayer in the presence of strong interactions $U_{c,I}\gg t_{c,I}$ can be mapped to an effective single-layer $t$-$J$ model. In order to perform the mapping, we consider $n_c=1$ and $n_I\leq 1$, and allow for sizeable $t_I\approx t_c$. Under these conditions, the distinction between conduction and impurity layers is no longer well-defined; nevertheless, we retain the original notation for the sake of consistency.

In this case, the low energy subspace consists of (bosonic) rung singlets
$$\Bd_{\mathbf{i}}=\frac{1}{\sqrt{2}}\left( \hat{c}_{c\mathbf{i}\uparrow}^\dagger\hat{c}_{I\mathbf{i}\downarrow}^\dagger-\hat{c}_{c\mathbf{i}\downarrow}^\dagger\hat{c}_{I\mathbf{i}\uparrow}^\dagger\right)$$ 
or rungs with an occupied $c$-site and empty impurity site, represented by new fermionic operators $$\hat{\Tilde{c}}^\dagger_{\mathbf{i},\sigma}=\Cd_{I\mathbf{i}\sigma} \B_{\mathbf{i}}.$$  
Since the particles in the half-filled $I$-layer are blocked due to the Gutzwiller single particle projection, only the impurity layer contributes to the in-plane hopping terms. This results in
\begin{equation}
\begin{aligned}
    \Ham_\mathrm{eff} = \mathcal{P}\left[ -\Tilde{t}\sum_{\langle \mathbf{i},\mathbf{j}\rangle}\hat{\Tilde{c}}^\dagger_{\mathbf{i}\sigma}\hat{\Tilde{c}}_{\mathbf{j}\sigma} + \hc \right]\mathcal{P}
    +  \Tilde{J} \sum_{\langle \mathbf{i},\mathbf{j} \rangle} \left[  \Tilde{\mathbf{S}}_\mathbf{i} \cdot \Tilde{\mathbf{S}}_\mathbf{j} - \frac{\gamma}{4}\hat{\Tilde{n}}_{i} \hat{\Tilde{n}}_{j}\right]+\frac{J_c}{2}\sum_{ \mathbf{i}}\Hat{\Tilde{n}}_\mathbf{i},
    \label{eq:Bilayer_to_tJ}
\end{aligned}
\end{equation}
with $\Tilde{t}=t_c/2$, $\Tilde{J}=J_I$ and $\gamma=\frac{J_c}{J_I}$. Note that in this effective single-layer $t$-$J$ description, creating a hole at site $\mathbf{i}$ corresponds to creating a singlet at rung $\mathbf{i}$ of the bilayer model. In the derivation of the mixD bilayer Eq.~\eqref{eq:mixDtJ}, an additional three-site term appears that is often neglected. This in turn gives rise to a three-site term in the effective single band model \cite{schloemer2024localcontrolmixeddimensions}.

\subsection{Weak interactions $U_{I}$: Mapping to a single-layer Fermi-Hubbard type model}

\begin{figure}[t]
\centering
\includegraphics[width=0.49\textwidth]{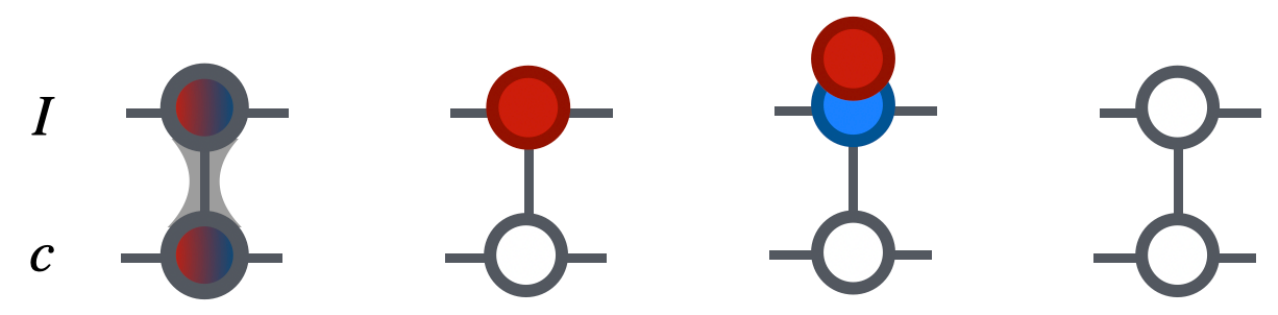}
\caption{ Low energy states for $J_K\gg t_c, t_I, U_I, U_c$. }
\label{fig:lowE}
\end{figure}
For smaller interactions $U_{I}$, the occupations in the $I$-layer are not constrained to single occupancy. In addition to the rung-singlets and singly-occupied rungs, we have two more low-energy configurations in this case that arise from doublon-hole fluctuations: rungs with empty sites for both layers and rungs with an empty $c$-site and a doublon at the $I$-site. For clarity, these low-energy configurations are schematically shown in Fig.~\ref{fig:lowE}.\\

In this case, we can rewrite the bilayer model Eq.~\eqref{eq:Kondo} (to first order) to an effective single-layer Fermi-Hubbard type model
\begin{align}
    \Ham_\mathrm{eff} = -\Tilde{t}\sum_{\langle \mathbf{i},\mathbf{j}\rangle}\mathcal{P}_S^{\mathbf{i},\mathbf{j}}\left[  \Bd_{\mathbf{i}}\B_{\mathbf{j}}\C_{I\mathbf{i}\sigma}\Cd_{I\mathbf{j}\sigma}  + \hc \right]\mathcal{P}_S^{\mathbf{i},\mathbf{j}}
    -t_I \sum_{\langle \mathbf{i},\mathbf{j}\rangle}\mathcal{P}_{\bar{S}}^{\mathbf{i},\mathbf{j}}\left[ \hat{{c}}^\dagger_{I\mathbf{i}\sigma}\hat{{c}}_{I\mathbf{j}\sigma} + \hc \right]\mathcal{P}_{\bar{S}}^{\mathbf{i},\mathbf{j}} +U_I \sum_{\mathbf{i}}{\n}_{I\mathbf{i}\uparrow}{\n}_{I\mathbf{i}\downarrow},
\end{align}
where $\Tilde{t}=t_c/2$, $\mathcal{P}_{{S}}^{\mathbf{i},\mathbf{j}}$ projects onto configurations with one rung-singlet at rung $\mathbf{i}$ or rung $\mathbf{j}$ and $\mathcal{P}_{\bar{S}}^{\mathbf{i},\mathbf{j}}$ projects onto no rung-singlet configurations at rungs $\mathbf{i},\mathbf{j}$. One can see that in this model, the hopping of the actual dopants (i.e. the singlets) in the first term is separated from the doublon-hole fluctuations in the second term. Furthermore, their relative strength can be tuned w.r.t. each other, which allows one to systematically study dopant dynamics and interactions with the environment.

\end{document}